\documentclass{PoS}

\usepackage{epsfig}
\usepackage{amsmath,epsfig,amssymb,dsfont}
\usepackage{amscd,latexsym}
\usepackage{graphicx}

\def\lsi{\raise0.3ex\hbox{$<$\kern-0.75em\raise-1.1ex\hbox{$\sim$}}}
\def\gsi{\raise0.3ex\hbox{$>$\kern-0.75em\raise-1.1ex\hbox{$\sim$}}}

\input macros.sty      
\input eqalign.sty     

\title{\hspace*{12.0cm} {\large \texttt{DESY 06-152}}\\
Status Report on ILDG activities}

\ShortTitle{Status Report on ILDG activities}

\author{Karl Jansen  \\
         John von Neumann-Institute for Computing NIC, \\             
         DESY, Platanenallee 6, D-15738 Zeuthen, Germany\\
E-mail: \email{Karl.Jansen@desy.de}}

\abstract{
A status report about the International Lattice Data Grid
(ILDG) is given. 
Different countries participating in the ILDG
have created regional lattice data grid solutions that are implemented,
working and used. The remaining task and the focus of present activities
is the development of the
interoperability of these regional grids.
A first, successful step in this direction is a 
metadata catalogue service which is already working 
interoperable.
}
\FullConference{XXIVth International Symposium on Lattice Field Theory\\
                July 23-28, 2006\\
                Tucson, Arizona, USA}

\begin{document}

\section{The ILDG idea and Participating Countries}

The wish to have a convenient tool to exchange valuable configurations from 
large scale lattice simulations is certainly high. Such a tool should 
\begin{itemize}
\item allow for a simple, semantic-based web search of 
configurations of interest;
\item enable the download of such configurations;
\item provide the configurations in a standardized format;
\item realize all this internationally and across borders.
\end{itemize}

The realization of such a wish became in sight 
with the rapid development of Grid technologies as used for international 
experiments such as the LHC. 
Inspired by a suggestion of R.~Kenway, 
Australia, Germany, Japan, UK and USA decided in an initial 
video conference organized by Edinburgh \cite{ildg1} 
to define a project of the 
{\em International Lattice Data Grid} (ILDG). Later, France and Italy  joined 
and, since 
the ILDG is an open project, 
it is to be expected and desired that additional countries will 
participate in the future, too.

Progress on defining ILDG standards has been presented during several
lattice conferences \cite{ildg-at-lat}.
It is the purpose of this write-up to show that ILDG is becoming
a real and working infrastructure which will support 
and hopefully ease further research
on lattice gauge theories.
Many of the questions raised
in the first meeting have not only been answered, 
but working solutions have been 
found and implemented. ILDG issues were discussed in half-yearly virtual 
video meetings which also served to observe and follow the 
progress of the project,
see the URL of ref.~\cite{ildg1} for a list of the 8 ILDG meetings 
that have taken place so far. The structure of ILDG is that it consists of
a board comprising representatives of each participating 
country\footnote{Present members are 
R.~Brower (USA), K.~Jansen (Germany, chairman), R.~Kenway (UK), 
D.~Leinweber (Australia), O.~Pene (France), L.~Tripiccione (Italy), 
A.~Ukawa (Japan). The chair is rotated yearly.}, 
a Metadata working group and a Middleware working group, the 
members of which will be listed below.
Let me also recommend the poster presentation at this conference 
of ref.~\cite{presentation} for 
additional information and more technical details. 

In order to set the frame, 
let us have a look at the (quite substantial) supercomputer 
resources that today can be 
used for 
lattice field theory (LFT) in the countries participating to ILDG:  
\begin{itemize}
\item Australia: The CSSM collaboration has access to about 2 TFlops%
\footnote{We give peak performances.} of compute power on
commercial machines installed in part at APAC (Canberra) and SAPAC (Adelaide).
\item France: The main resource is apeNEXT with 1.2 Tflops, 
installed at the university 
Rome I ``La Sapienza''.  
\item Germany: There are 6 Tflops apeNEXT systems, installed in Bielefeld and in 
Zeuthen. In addition, there is (peer reviewed) computer time available at the 
German national supercomputer centers, 45 Tflops of a BG/L system and a 10 Tflops
IBM Regatta system at the 
Research center J\"ulich and a 26 TFlops Altix System at the LRZ in Munich. 
Lattice physicists
typically have access to about 10--20\% of this computer power.
\item Italy: Again, the working horse is a 7.2 Tflops apeNEXT system, installed 
at the university Rome I ``La Sapienza''. 
\item Japan: Here the machines are a 14.3 Tflops PACS-CS cluster in Tsukuba and a
57.3 Tflops BG/L system at KEK. In addition, there are smaller 
O(1) Tflops installations
in Hiroshima, KEK and Kyoto.  
\item United Kingdom: The major source of computer power is a 
12 Tflops QCDOC system,
installed in Edinburgh. 
\item USA: In the US, there is the 12 Tflops QCDOC machine in Columbia,
two 10 TFlops QCDOC machines at BNL,
and a total of about 10 Tflops cluster systems at Fermilab and JLAB. 
In addition, peer reviewed computer time is available at national supercomputer 
centers at NERSC, ORNL and Pittsburg. 
\end{itemize} 

In most of the countries there are ambitious plans for future increase of 
computer power with the aim to reach Petaflops computing soon.

When we add up the above listed resources, we find a total of about 
150 Tflops for LFT today.
The available computer time clearly allows for a significant push of the
simulation parameters towards a physical situation with small enough 
pseudo scalar masses, large enough physical volumes and small 
lattice spacings. 
The configurations generated are very 
precious and it is one of the aims of the ILDG to make best use of 
these configurations.
Already now, large collaborations 
-- UKQCD, RBC, JLQCD, QCDSF, MILC, CSSM, CP-PACS/PACS-CS, ETMC, 
SESAM/T$\chi$L/GRAL --
are storing configurations on the grid, employing ILDG standard format, 
using ILDG infrastructures and making these configurations available 
to the corresponding members of the collaborations for further analysis. 

The upload of configurations is proceeding rapidly and today we can 
already find more than 70.000 dynamical configurations residing in ILDG and 
waiting for download. In table~\ref{confs} I summarize those physics plans
of various collaborations from which configurations will be put on the 
grid. As can be seen, a great variety of actions is used and the set of 
configurations that will become available eventually is certainly 
very interesting. 
Besides the configurations listed in table~\ref{confs}, there exist  
configurations from older simulations, i.e. staggered $N_f=2$ from MILC, 
Wilson $N_f=2$ from SESAM/T$\chi$L/GRAL and $N_f=2$ tadpole improved Wilson
from CP-PACS.

\begin{table}
\begin{tabular}{llllll}
\hline
Collaboration & fermion action & Flavors & $a$[fm] & 
$am_q/m_\mathrm{PS}[GeV]$ & $aV_\mathrm{max}$ \\
\hline
\\
RBC/UKQCD & domain wall  & $N_f=2+1$ & 0.12 & $ m_q=0.0[2,3,4]$ & $24^3\cdot 48$ \\
MILC & rooted staggered  & $N_f=2+1$ & 0.06-0.125 & $\frac{m_l}{m_s}=0.[1,2,3]$ & $48^3\cdot 144$ \\
PACS-CS & NI Wilson & $N_f=2+1$ & 0.07-0.12 &  --- & $28^3\cdot 56$ \\
CSSM & FLIC & $N_f=2+1$ & 0.12 & $ m_\mathrm{PS}=0.3$ & $20^3\cdot 40$ \\
QCDSF & NI Wilson & $N_f=2$ & 0.05-0.11 & $ m_\mathrm{PS}=0.25-1$ & $32^3\cdot 64$ \\
ETMC & tm Wilson & $N_f=2$ & 0.07-0.12 & $ m_\mathrm{PS}=0.25-0.5$ & $32^3\cdot 64$\\
\hline
\end{tabular}
\label{confs}
\caption{Configurations that are or will be in the near future on the grid. 
Quark masses and volumes are given in lattice units. $am_l$ denotes the light, 
$am_s$ 
the strange quark mass.
NI stands for non-perturbatively improved and tm for (maximally) twisted mass 
fermions.
$aV_\mathrm{max}$ denotes the maximal lattice size that is planned for the 
simulations. In general, a sequence of also smaller lattice sizes is aimed 
for. Note that for the non-perturbatively improved 
Wilson fermion simulation the Iwasaki gauge action (PACS-CS) and the 
Wilson gauge action (QCDSF) will be employed.
}
\end{table}

The different collaborations have setup their own corresponding policy 
for allowing
to download and 
use these configurations. Possible rules are {\em immediate access; 
an acknowledgment in papers that use these configurations; waiting periods 
of six months between upload of configurations and giving access to them; 
draft of papers using the configurations in advance or a waiting period 
for the submission of key publications}. In addition, most collaborations 
want {\em citations of their key work for which these configurations were 
originally produced}. It appears therefore to be a wise idea to contact
the collaborations before accessing the configurations 
and ask for their particular policy. Some of them
are also thinking of collaborating on certain physics questions. 

It should be emphasized that uploading configurations is  
non-trivial and needs quite some work. I believe therefore that 
there should be an 
applause to those
collaborations who are willing to spend such an effort and an appeal 
to other collaborations to follow these examples. 

\subsection{Finding Configurations}

\begin{figure}
\begin{center}
\includegraphics[width=15.7cm,height=15.7cm]{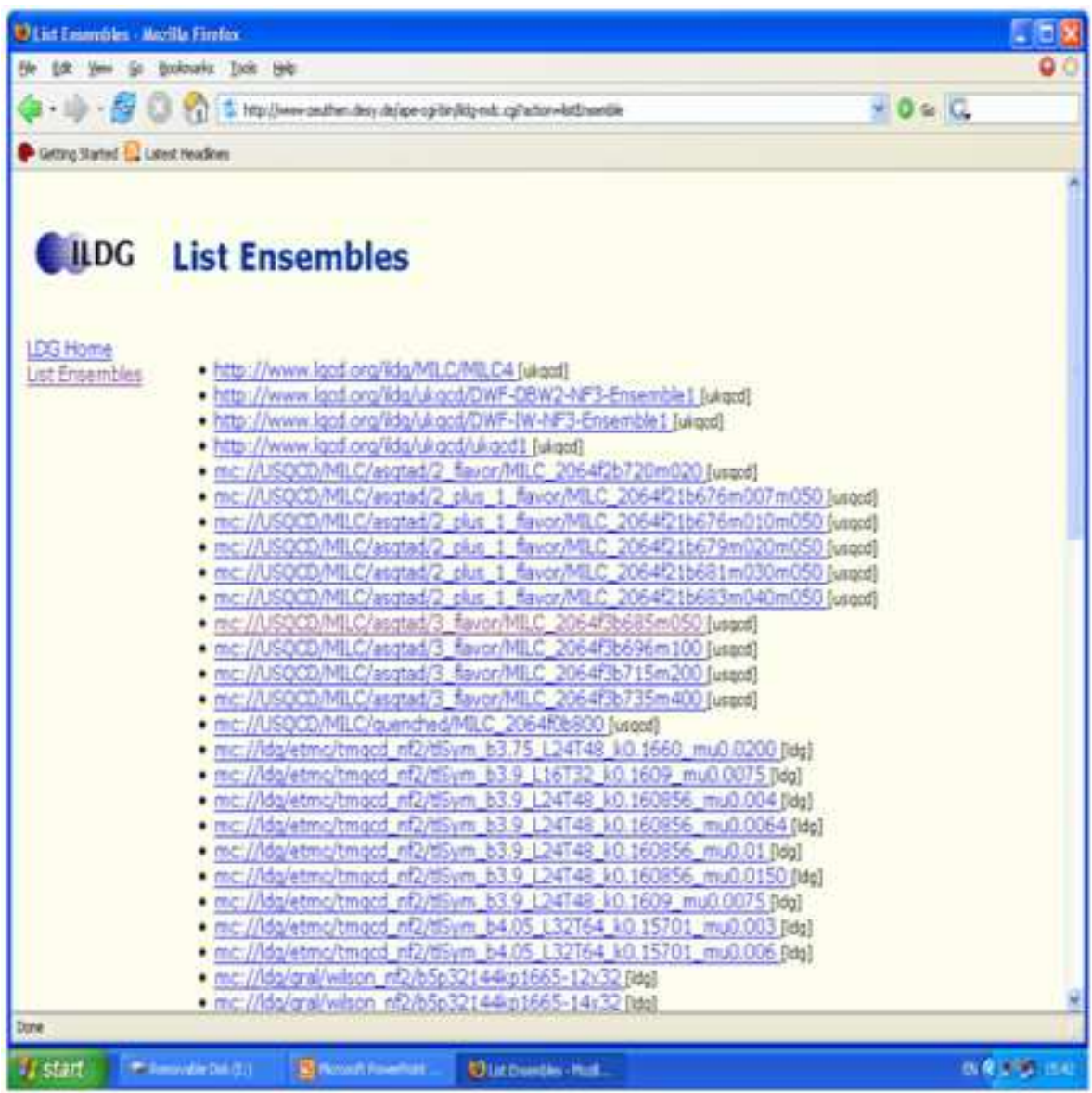}
\vspace*{-0.0cm}
\caption{The list of uploaded ensembles of configurations as seen when
using the LDG metadata catalogue.} 
\label{ensemblelist}
\end{center}
\end{figure}

\begin{figure}
\begin{center}
\includegraphics[width=15.7cm,height=15.7cm]{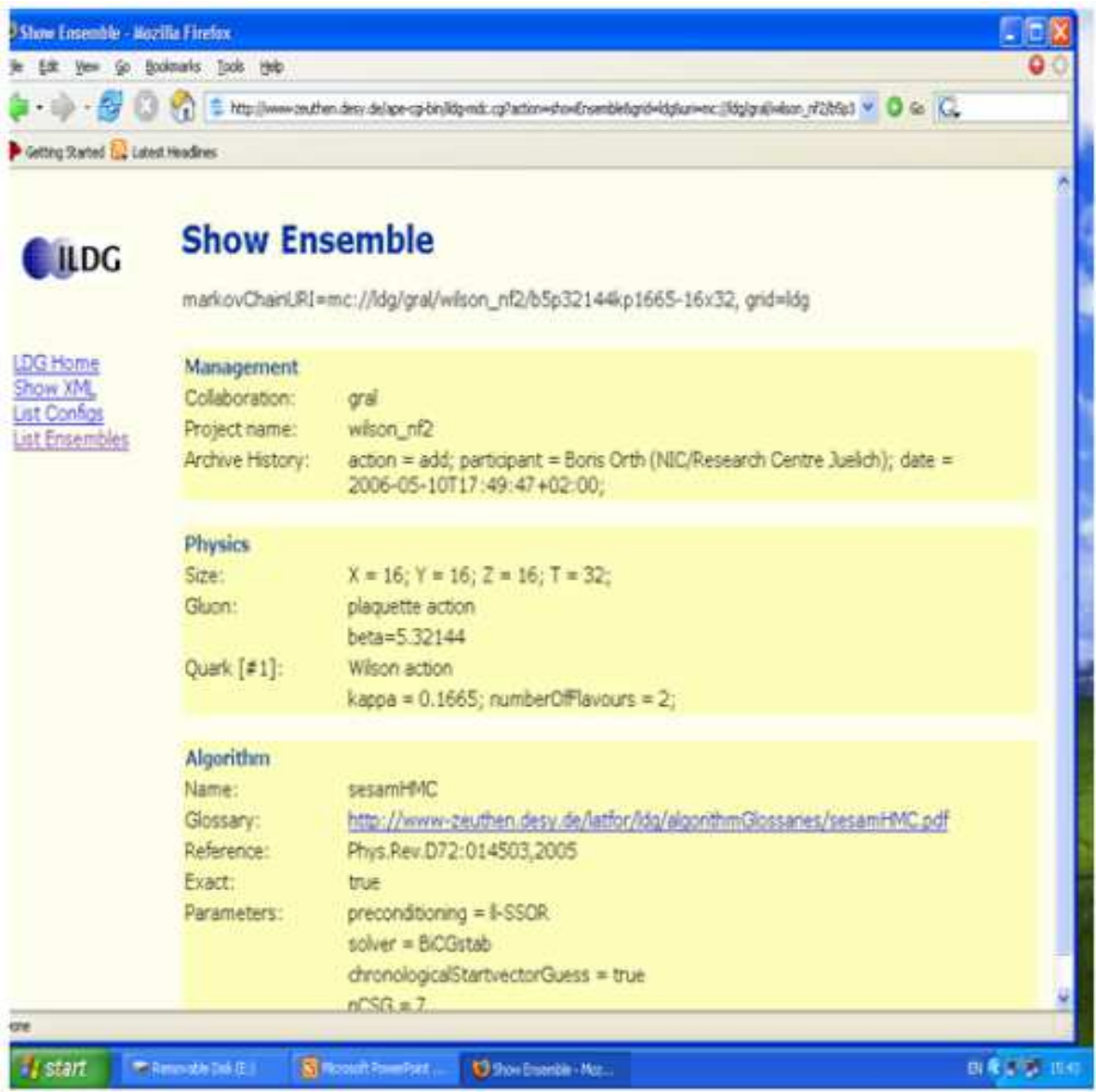}
\vspace*{-0.0cm}
\caption{The list of configurations when one of the ensembles is selected.}
\label{configurationlist}
\end{center}
\end{figure}

In order to find the configurations listed in table~\ref{confs}, you  
have to query the various metadata catalogues of the regional grids.  For
those catalogues which are already ILDG compliant, one of the web interfaces
which act as a portal and allow to query all catalogues can be used.
Let us take the one operated by the German/French/Italian LatFor Data Grid
(LDG) \cite{ldg} as an example. Going to 
the site given in \cite{ldg}, you will find a {\em list of ensembles}, 
see fig.~\ref{ensemblelist}. From the list of ensembles you can select 
a {\em list of configurations}, see fig.~\ref{configurationlist}. 
Finally, you can select a particular configuration to obtain the 
specific information for this configuration in a (hopefully)
self-explanatory manner, see fig.~\ref{configurationinfo}. 
To actually retrieve configuration brings us to the next section.

\section{Getting Configurations} 

\begin{figure}
\begin{center}
\includegraphics[width=15.7cm,height=15.7cm]{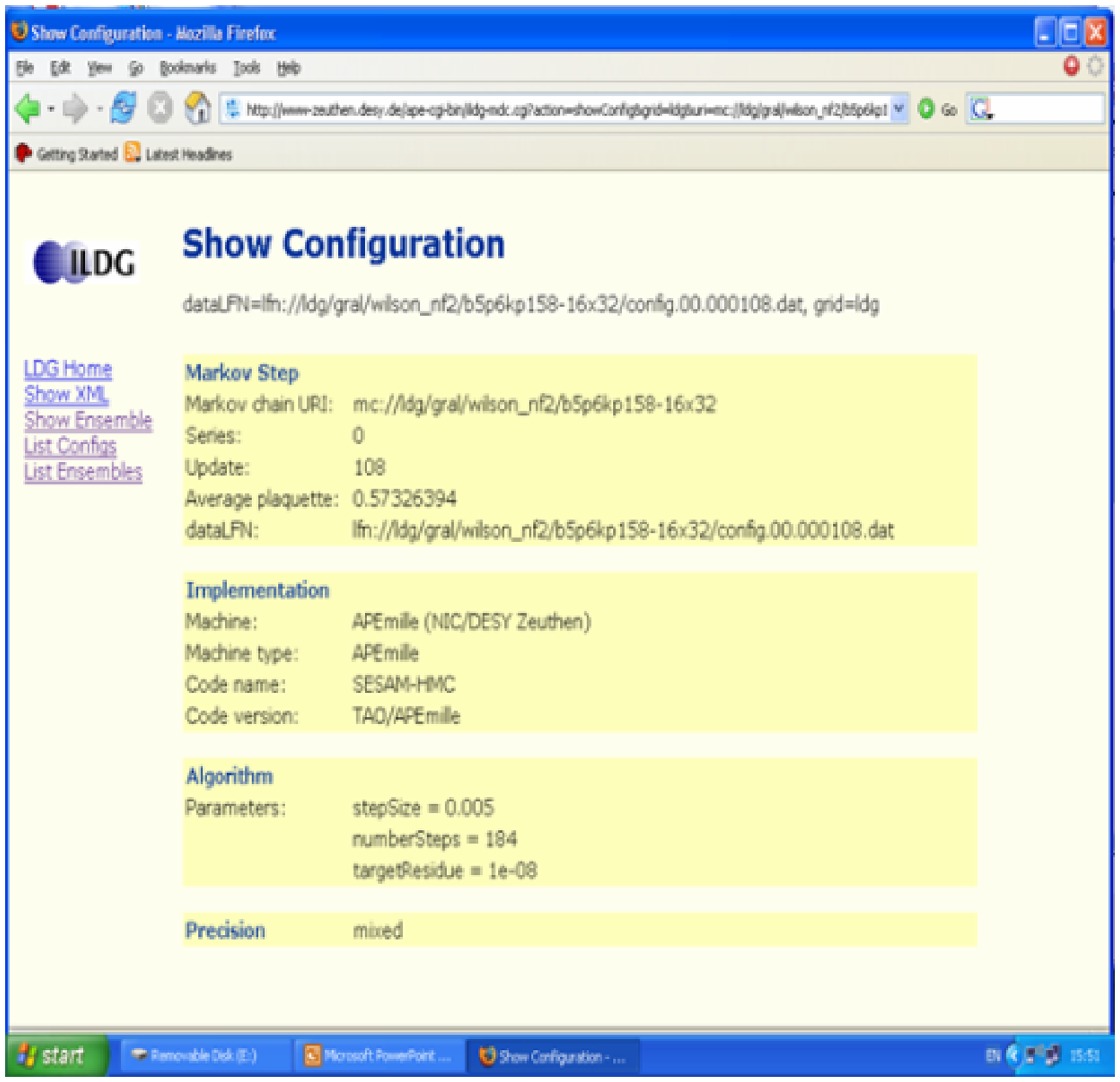}
\vspace*{-0.0cm}
\caption{The information on a selected configuration.}
\label{configurationinfo}
\end{center}
\end{figure}

Let us assume that you have browsed the metadata catalogue on the web 
as explained above,
and you found your favorite set of configurations that just fits to address 
your physics problem. 
Let us further assume that you have even 
contacted the corresponding collaboration and you got green light 
for downloading their configurations. Here are then the next steps to proceed: 

\begin{itemize}
\item The first thing you have to do is to get a grid certificate. For this
you have to identify a local Certificate Authority (CA), which is willing
to provide you with a certificate. Before you receive a certificate, the CA
or one of its Registration Authorities (RA) will check your identity,
e.g.~ask for your passport.
It is foreseen that ILDG resource providers will trust certificates
issued by any CA that is member of the International Grid Trust Federation
(IGTF, \cite{igtf}).
Of course, the step of obtaining a grid certificate has to be done 
only once.
\item As a next step you have to become member of the Virtual Organisation
(VO) ILDG. There is a policy 
being worked out about who can become 
a member of the VO ILDG. Basically, this is open for all people doing 
lattice field theory. Each regional grid has to nominate two representatives 
who can decide whether a particular person shall become a member of the 
VO ILDG.
\item The next thing you have to do is to install software that will allow 
you to actually download the configurations.
This software will depend on the particular regional implementation of 
ILDG in your country (see also the discussion on regional grid setups below). 
Let me take, for some convenient reason, as an example again 
the LatFor Data Grid (LDG) \cite{ldg}. 
At our site, you would retrieve a Grid User Interface and
a package with user tools called \texttt{ltools} \cite{ltools}
which runs on basically all Linux platforms. 
\item After you have obtained your grid certificate and you have 
installed the software
you are now ready to get your favorite set of configurations. 
Let us take as an example the set you find on the web as shown in 
fig.~\ref{configurationinfo}. What you will detect is a 
{\em Logical File Name} (LFN) for the configuration you want to obtain. 
This LFN represents the globally unique grid address of your configuration and issuing
\texttt{lget lfn} will then get you the desired configuration 
directly on your disk (if you have made sure that you have enough space there).
Naturally, if you want a whole set of configurations, you will write a 
small script to retrieve all desired configurations.
\end{itemize}

\vspace*{-0.0cm}

\vspace*{-0.0cm}

\vspace*{-0.0cm}

It could be asked whether this all is worth the effort. 
On the negative side, there is certainly the initial task of 
getting your grid certificate and installing the software to make 
use of ILDG on your regional grid. But then, you can download very 
precious configurations on which you can address your physics 
application (which should, of course, not be in conflict with the configuration 
owner's ideas). Moreover, you know exactly the format in which the 
configuration is written and you are sure that for future downloads 
this format will be the same. Therefore, your measurement code will run
immediately on any new set of configurations. 
As an additional point, it should be stressed that once you have overcome the
initial difficulties, the download of configurations will be routine.
Thus, the ILDG mechanism is a new way to use
configuration for all their potential of physics applications, avoiding
duplication and loss of information - something that has been very 
common in the
past - and therefore increasing the overall efficiency of the community.

Of course, it remains to be seen in practice, whether this optimistic 
point of view is indeed realized or whether we are left with some
practical issues that will drive life complicated. 

Uploading the configurations is more tricky and needs extra work. 
First of all, configurations have to be stored in the ILDG configuration format, see
ref.~\cite{ildg1}, a Logical File Name (LFN) has to be added and the metadata 
have to be created. For doing all this, the 
freely available LIME library is needed which can be obtained 
from the URL in ref.~\cite{lime}.
The metadata consist of the {\em ensemble XML file}, 
the {\em configuration XML file},
a {\em glossary file} and a {\em configuration checksum}. 
Finally, you have to store the binary configuration into a storage element
and the metadata into the metadata catalogue.
This typically can be done using just a single command. For instance,
in LDG this command is called 
\texttt{lput}, where you have to specify the 
storage element you actually
want to store the file physically and the names of the
metadata document as well as the file containing the configuration.

Although there are many help tools on the corresponding ILDG sites, and scripts exist
to automatize the upload procedure, there is clearly some effort necessary 
to store the configurations.

\section{Looking behind the scene}

The fact that we have today solutions and working implementations of 
regional grids realizing the ILDG idea, 
is purely due to the hard work of people
involved in the metadata and the middleware groups listed here
\footnote{In addition, there are many people working together with 
the Metadata and Middleware working groups. The names of these people
can be found on the webpages of the corresponding collaborations.}:
\begin{itemize}
\item Metadata Working Group: G. Andronico, P. Coddington, 
C. DeTar, R. Edwards,
B. Joo, C. Maynard, D. Pleiter, J. Simone, T. Yoshie
\item Middleware Working Group: G. Beckett, D. Byrne, M.Ernst, B. Joo,
D. Pleiter, M. Sato, C. Watson
\end{itemize}

\begin{figure}
\begin{center}
\includegraphics[width=12.8cm,height=12.7cm]{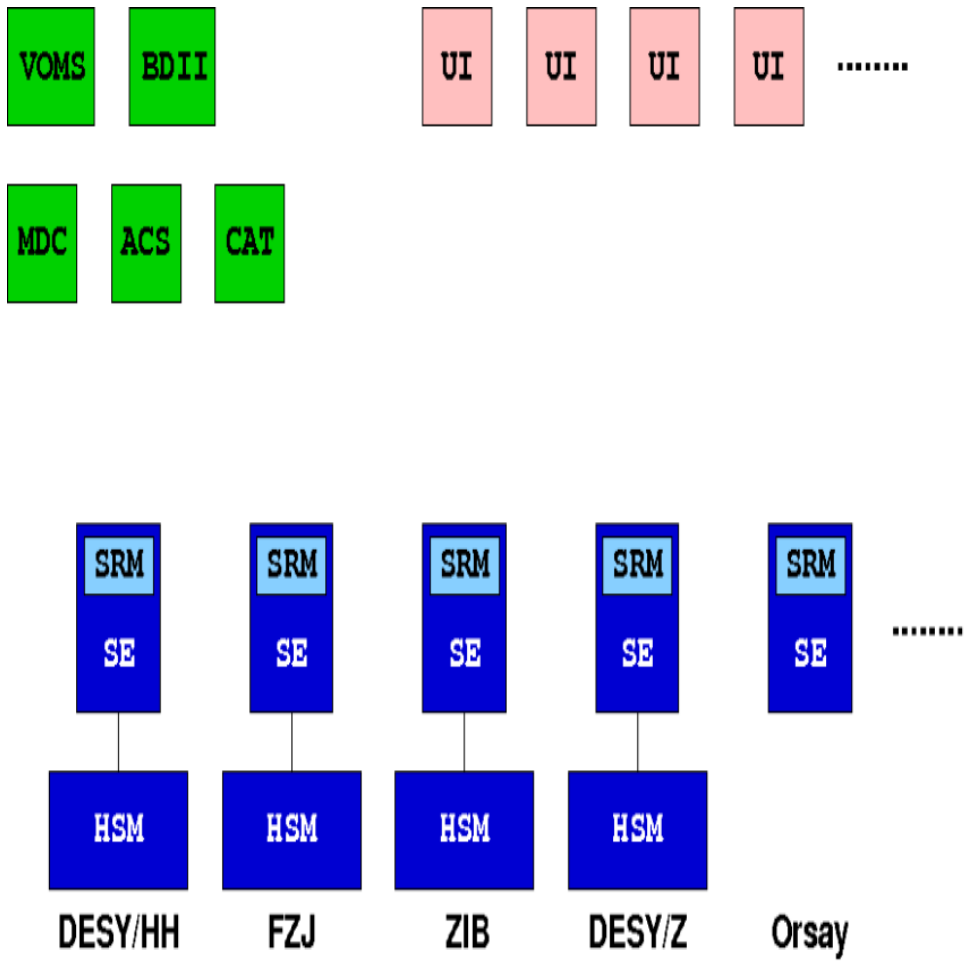}
\vspace*{-0.0cm}
\caption{The major components for a grid infrastructure. The abbreviations 
are: UI: User Interface, SE: Storage Element, SRM: Storage Resource 
Manager, VOMS: Virtual Organization Management System, MDC: Metadata
Catalogue, BDII: Berkeley Database Information Index, 
HSM: Hierarchal Storage Management, ACS: Access Control Service, CAT:
File Catalogue.
All of these components have to work together smoothly in order to have 
a functioning regional grid infrastructure.}
\label{components}
\end{center}
\end{figure}

In order to establish even a regional grid infrastructure is 
highly non-trivial. 
Many components have to work together as sketched in 
fig.~\ref{components}. It is 
therefore a very big step forward that at basically all participating sites 
regional grid infrastructures have been developed. 
Let me list the characteristics of the different regional grids that 
exist today.

\vspace*{-0.0cm}

\begin{itemize}
\item Germany/France/Italy \\
These countries use the LatFor data grid (LDG).
LDG has a metadata catalogue \cite{ldg-mdc} to
query, download, upload and manage metadata.
The user software tool is \texttt{ltools} with commands such as
\texttt{lget, lput, lls, lupdate} to download, upload, list and update
configurations, respectively. The regional grid infrastructure is based
on LCG-2 compliant components. For further details see ref.~\cite{ldg}.
To store configurations, there are 50 Terabytes storage space in Germany and
5 Terabytes in France available.
\item United Kingdom\\
In the UK the QCDgrid is used.
QCDgrid also has a metadata catalogue based on the native XML-database
``eXist'' \cite{exist} complete and deployed on the UKQCD
development system. It is now running on the production system 
providing access to the real metadata. 
Access is realized through the ILDG sample
clients metadata.
User tools come, as within LDG, as command line tools such as
\texttt{put-file-on-qcdgrid, get-file-from-qcdgrid}. In addition,
the UK QCDgrid is
working on developing web-based tools that will allow for
graphical user interfaces to
handle the metadata. The storage space in the UK is about
80 Terabytes across the UK
but mainly in Edinburgh, see ref.~\cite{qcdgrid}.
\item Japan\\
The Japanese Lattice Data Grid (JLDG) has also implemented an
eXist-based metadata catalogue.
The user software consists of commands such as \texttt{Gftp} on a specially
developed Gfarm file system. Security is realized by the
Grid Security Infrastructure (GSI) \cite{gsi}.
Also in Japan 50 Terabytes of storage space is available for the ILDG.
More information
is provided at ref.~\cite{jldg}.
\item USA\\
The USQCD has a metadata catalogue up and running.
The metadata catalogue is constructed as a web service. The user software
to download and
upload configurations is not yet completely finished.
USQCD has 50 Terabytes
of storage elements at NERSC, BNL and FNAL.
Further information can be found at ref.~\cite{usqcd}.
\item Australia\\
In Australia we again see the metadata catalogue to be complete.
User software for handling configurations is realized through a web portal.
In Australia we find 25 Terabytes for storage.
Additional information is given at the ref.~\cite{cssm}.
\end{itemize}

The above list demonstrates that indeed substantial progress has been achieved 
towards the implementation of the ILDG idea:

\begin{itemize}
\item We have five working regional ILDG infrastructures ready, 
LDG (Germany/France/Italy), JLDG (Japan),
QCDgrid (UK), USQCD (USA) and Australia. 
\item There are about 250 Terabytes of storage space available at these sites. 
This would allow to store, roughly, 200.000 configurations on a 
$32^3\cdot 64$ lattice which 
would need some time to be generated. 
\item Actually, already now 70 000 configurations are available on 
the ILDG (many of them 
on smaller lattices than $32^3\cdot 64$).
\end{itemize}

\section{Interoperability} 

\begin{figure}(0,58)
\put(-10,10){\includegraphics[width=7.0cm,height=7.0cm]{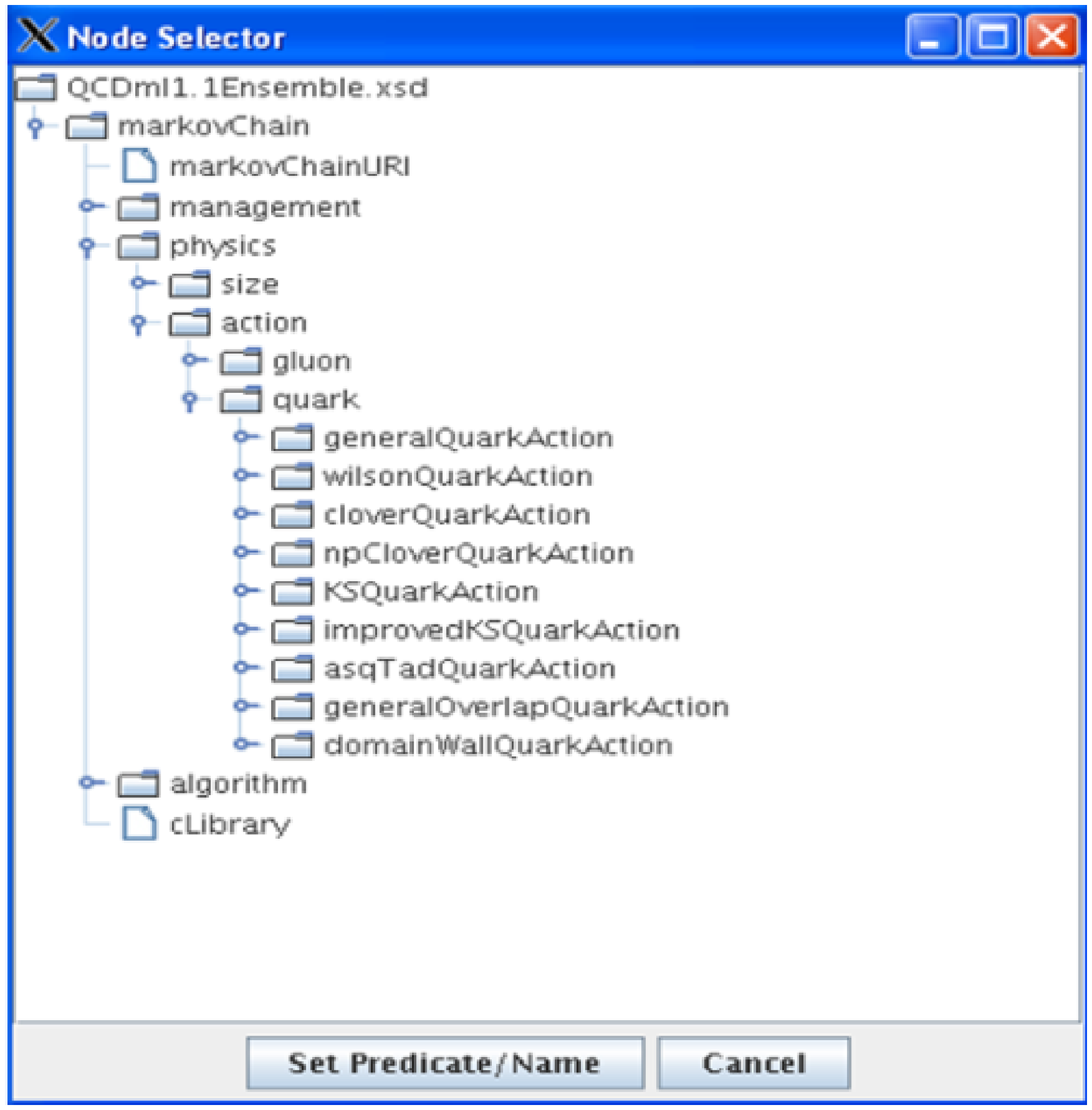}}
\put(70,10){\includegraphics[width=7.0cm,height=7.0cm]{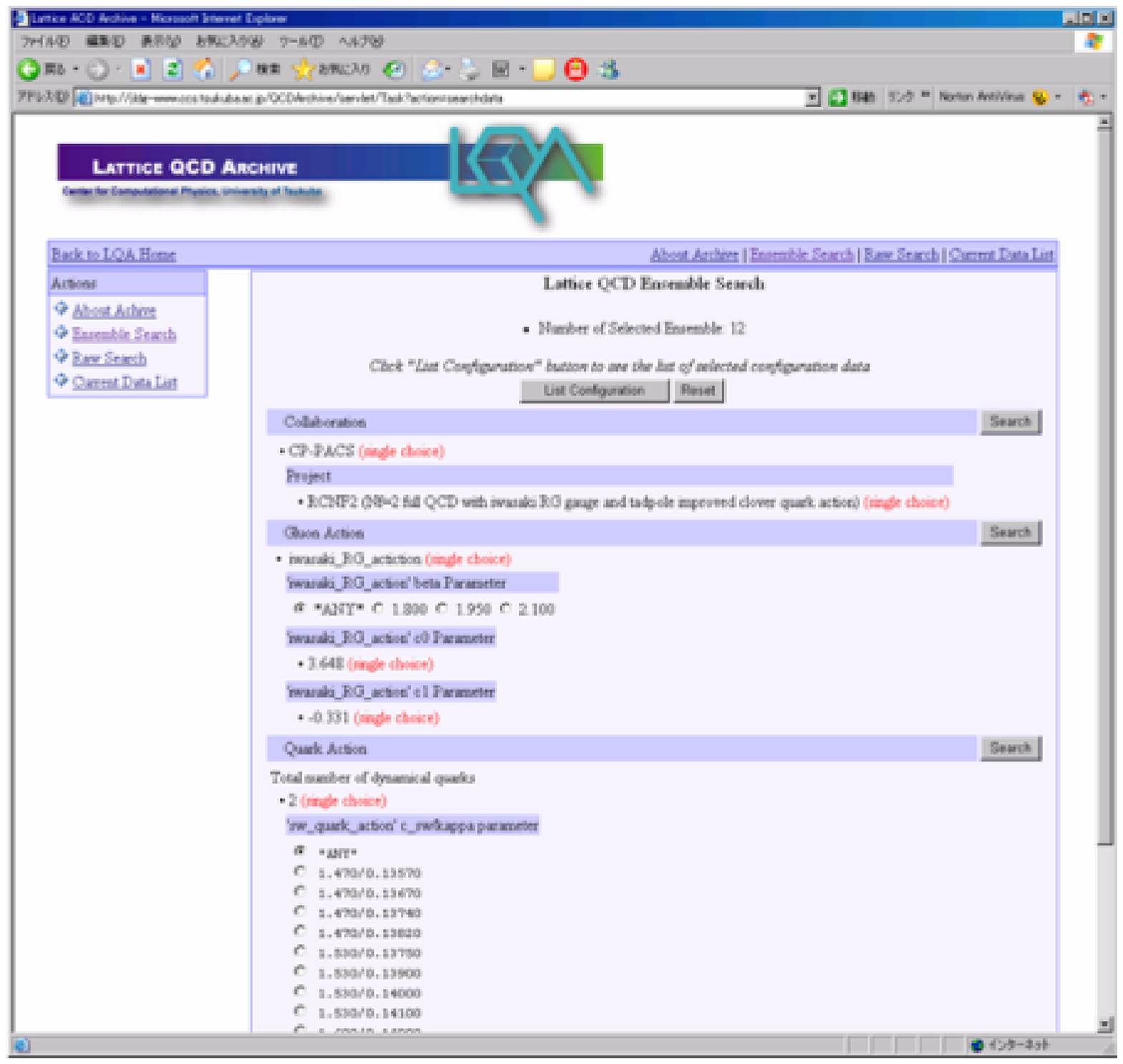}}
\put(-10,-65){\includegraphics[width=7.0cm,height=7.0cm]{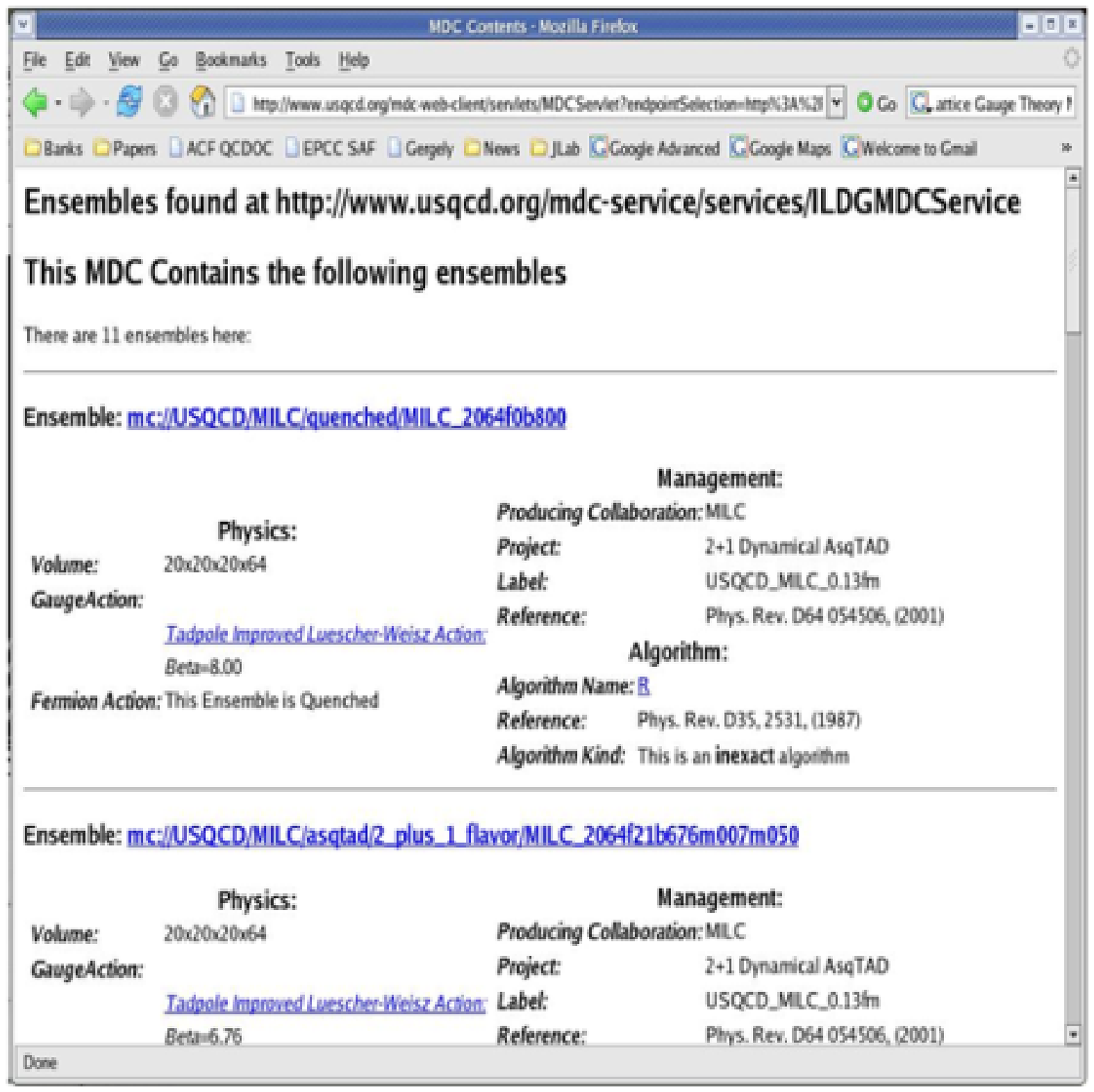}}
\put(70,-65){\includegraphics[width=7.0cm,height=7.0cm]{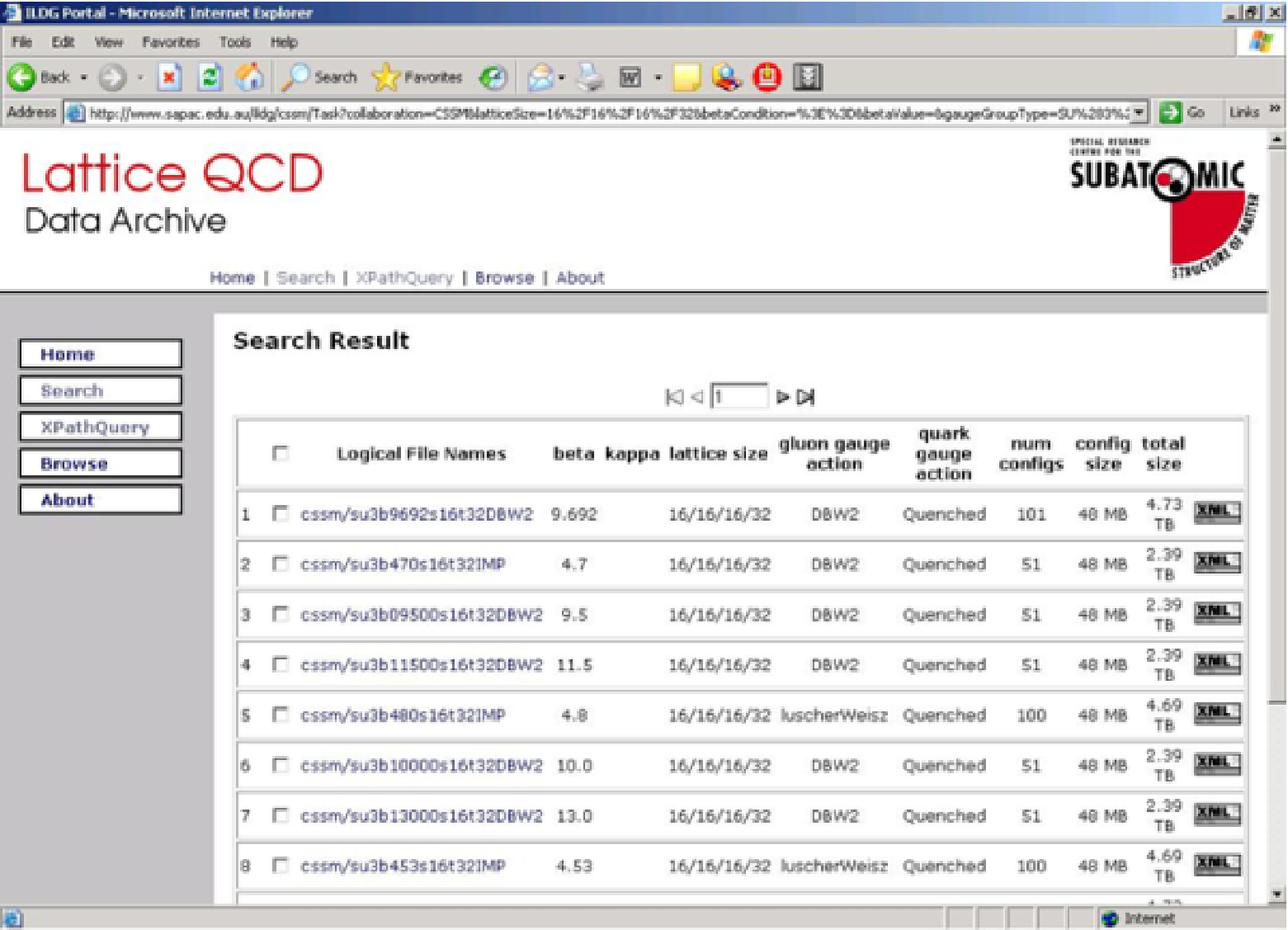}}
\caption{The appearance of different web browsers for listing configurations:
UK QCDgrid (upper left), Japan JLDG (upper right), USA USQCD (lower left) 
and Australia (lower right). For examples of the German/France/Italy LatFor 
Data Grid (LDG) see 
figs.~\protect\ref{ensemblelist}-\protect\ref{configurationinfo}.}
\label{allmeta}
\end{figure}

With the regional grids now functioning, ILDG has to face the next challenge, 
namely the interoperability of these regional grids. One big progress in this 
context is the operation of inter-operable metadata catalogue services. 
This means that it is possible for the sites to browse all each other's 
metadata catalogue, see fig.~\ref{allmeta} for examples of the 
appearance of the web-browser at the participating sites. 
In the development of this service, it was necessary to define the interface
in terms of the web service description language (WSDL) and to agree
on a behavioral specification.%
\footnote{WSDL essentially defines the name of the services as well as
the name and type of the input arguments and the return values.
The behavioral specification defines, e.g., what should happen
in case of errors, 
which status codes are returned, etc.}
In addition, a set of tests has been defined
to verify ILDG compliance of particular services.

\vspace*{-0.0cm}
\setlength{\unitlength}{1mm}

However, there are still most significant efforts required to achieve the 
interoperability of remaining components which are: 
{\em Security, File Catalogues and
Storage Elements}. It is worth stressing that also here progress has been made
already. It could be demonstrated successfully that file transfers between 
LDG (at DESY), QCDgrid (at EPCC) and USQCD (at Fermilab and JLAB) are possible. 
Another point is that through 
a virtual organization membership service (VOMS) provided by the global grid 
community the VO ILDG can be managed. A draft for policies on who can become a 
member of the 
VO ILDG exists and will be completed soon.

\section{Summary} 

In discussions with many lattice practitioners, it was clearly expressed 
that the ILDG idea is extremely useful and valuable. Nevertheless, there was 
always a number of questions, typical ones (with some answers) are:

\noindent {\em Can I determine access rights myself?}\\
This very desirable feature is currently only supported at the regional 
grid level which allow to define groups, rights for each group and ensembles of
configurations. 
In particular, it is not only possible to manage access rights for 
configurations,
but to also delete and replace them. 

\noindent {\em Will the data be replicated?}\\
Currently, replication is possible only within the regional grids. 
It is, however, planned to make replication beyond grid boundaries possible. 

\noindent {\em How can I check that I got the right configuration?}\\
For each configuration a plaquette value and a checksum is provided. 

\noindent {\em Is the schema to describe data extensible?}\\
Yes, more information can always be built in.

\noindent {\em Can I put in algorithm information?}\\
It is possible to add a name space for algorithmic information. However, ILDG 
demands only a limited information by default since algorithmic parameters are 
plenty and it will become too complicated to manage them. 

\noindent {\em What about propagators?}\\
The ILDG is presently evaluating the possibility to have a 
simplified version, e.g. 
standardized format, for propagator storing. In general, it appears to be too 
complicated to repeat the example of configurations but discussions 
are ongoing. 

\noindent {\em Who is paying for all this?}\\
In most countries, ILDG is embedded in larger grid projects that are mainly
oriented towards LHC or other large scale experiments. ILDG profits from these 
developments and makes use of the software and hardware infrastructure that 
becomes available through these projects. For example, EPCC at Edinburgh
and DESY at Hamburg/Zeuthen are Tier 2 centers within the LHC grid (LCG).
In addition, ILDG is involved in a number of 
grid projects financed on a national or European level.

In conclusion, much progress has been made to develop regional grids for the 
ILDG idea. These infrastructures are presently successfully used by large
collaborations, which have members at many different sites,
to exchange their configurations. Again it should be stressed that the 
realization of these regional grids do not come for free but through the 
hard work of people involved in building-up these infrastructures.

The next milestone and target has to be the interoperability of 
these regional grid
solutions. Here, the possibility of browsing each other's metadata 
catalogue has been successfully 
demonstrated, already. Even file transfer between several 
sites (DESY, EPCC, Fermilab/JLAB) has been achieved.
Both of these accomplishments are important steps towards interoperability.

Thus, it remains to emphasize that it is time now to get a grid certificate at 
your local authorization site and become a member of the virtual organization 
ILDG.

\section{Acknowledgments}
This presentation at the 24th International Symposium of Lattice Field 
Theory in Tucson
would not have been possible without the help of people sending me the material 
of their corresponding country. I am very much indebted to 
R.~Brower, D.~Byrne, B.~Coddington, B.~Joo, R.~Kenway, D.~Leinweber, 
C.~Maynard, O.~Pene, D.~Pleiter, M.~Sato, 
L.~Tripiccione, 
A.~Ukawa, A.~Williams, T.~Yoshie. 
A special thanks goes to D.~Pleiter for a critical reading and very useful 
comments to this manuscript. 
Finally, in the name of all ILDG members, I would like to thank the 
organizers of the 
conference to give the room to present the status of the ILDG.


\end{document}